\documentclass{article}
\usepackage{spconf,amsmath,amssymb,graphicx}
\usepackage{xcolor}
\usepackage{makecell}
\usepackage{pdfpages}
\usepackage[]{algorithm2e}
\usepackage{todonotes}
\usepackage{float}
\usepackage{calc}
\usepackage{graphicx}
\usepackage{color}
\usepackage{subcaption}
\usepackage{balance}
\usepackage[final]{changes}
\usepackage{pgfplots}
\usepackage[hyphens]{url}
\usepackage{changes}

\ninept

\DeclareMathOperator{\logit}{logit}

\DeclareMathOperator{\diag}{diag}
\DeclareMathOperator{\llr}{llr}

\DeclareMathOperator*{\argmax}{argmax}
\newcommand{\Dece}{D_{\text{ECE}}}
\newcommand{\Cllrmin}{C_{\text{llr}}^{\text{min}}}

\title{A bridge between features and evidence\\for binary attribute-driven perfect privacy}
\name{Paul-Gauthier Noé$^1$, Andreas Nautsch$^2$, Driss Matrouf$^1$, Pierre-Michel Bousquet$^1$, Jean-François Bonastre$^1$}
\address{$^1$Laboratoire Informatique d'Avignon, Avignon Université, France\\
  $^2$vitas.ai, Germany}
\begin{document}
%
\maketitle
\begin{abstract}

Attribute-driven privacy aims to conceal a single user's attribute, contrary to anonymisation that tries to hide the full identity of the user in some data. When the attribute to protect from malicious inferences is binary, perfect privacy requires the log-likelihood-ratio to be zero resulting in no \emph{strength-of-evidence}. This work presents an approach based on normalizing flow that maps a feature vector into a latent space where the \replaced{\emph{evidence}}{strength-of-evidence}, related to the binary attribute, and an independent residual are disentangled. It can be seen as a non-linear discriminant analysis where the mapping is invertible allowing generation by mapping the latent variable back to the original space. This framework allows to manipulate the log-likelihood-ratio of the data and therefore allows to set it to zero for privacy. We show the applicability of the approach on an attribute-driven privacy task where the sex information is removed from speaker embeddings. Results on VoxCeleb2 dataset show the efficiency of the method that outperforms in terms of privacy and utility our previous experiments based on adversarial disentanglement.

\end{abstract}
\section{Introduction}
\label{sec:intro}

Attribute-driven privacy aims to make the data independent of an attribute a user wants to keep secret~\cite{noe21_interspeech,aloufi2020privacy}. In~\cite{noe21_interspeech} we proposed a method based on adversarial disentanglement where the data is fed into an autoencoder and where there is an attribute's class classifier on the encoded representation. With the adversarial training, the encoder tries to fool the classifier and the latter tries to predict the right attribute's class. This reduces the discriminant information related to the attribute in the encoded representation. However, the way this approach protects the data may remain obscure and explainability may be jeopardised. In this paper, we explore a new vision of binary attribute-driven privacy inspired by probabilistic forensic science. Indeed, understanding how the attribute's information is embedded in the data helps in designing a protection system. Even if the term \emph{evidence} is used sometimes to denote the observed data, we use evidence here to point what is useful for the decision task in the observation~\cite{meester_slooten_2021}. Also, we call \emph{residual} anything in the observation that is independent of the variable (attribute) of interest. Therefore, our aim is to disentangle the evidence and the residual from the observation. The \replaced{evidence}{\emph{strength-of-evidence}} is represented by a likelihood-ratio~\cite{CHAMPOD2000193} between two mutually exclusive propositions which correspond here to the two classes of the attribute. For perfect privacy~\cite{Nautsch2020}, the log-likelihood-ratio (LLR) must be zero to alleviate the \emph{strength-of-evidence} of the data: this is zero-evidence. \added{As far as we know, there is no literature in the speech community that deals with privacy with the terms of \emph{perfect privacy}~\cite{NAUTSCH2019441,Tomashenko2020} except~\cite{Nautsch2020} but it presents a privacy evaluation rather than a protection method. }\replaced{Even so, L}{L}inear Discriminant Analysis (LDA)~\cite{murphy2012machine} allows to linearly map the data into a space where the discriminant component lives in a logit posterior line on which the LLR can be identified. However, the Gaussian and shared covariance assumption of the LDA and the resulting linear mapping may result in a poor estimation of the LLR and is not flexible enough for real data. Therefore, we propose a non-linear discriminant analysis that maps the data into a latent space where the class-conditional densities are Gaussian with same covariance matrix. The first component is the only discriminant and is a LLR while the other components are the residual variabilities irrelevant for the attribute inference. Moreover, this mapping needs to be invertible in order to allow the manipulation of the evidence inside the latent space and the reconstruction into the original feature space\footnote{Quadratic discriminant analysis is an example of a non-linear data transformation onto a logit posterior space. However, it is still based on a Gaussian assumptions and it does not allow a return into the feature space~\cite{bishop,dimredhastie}.}.
In order to do so, the proposed method is based on Normalising Flow (NF)~\cite{normflowkobyzev,dinh2017}. This is a generative model that learns a bijective mapping between the feature and the latent space. Most of the time, it is used in an unsupervised framework where the latent distribution is a multivariate Gaussian distribution. Some works proposed to use NF in a supervised case~\cite{izmailov2020semi,atanov2019semi}. In our case, the two class-conditional densities are mapped into two multivariate Gaussian distributions where their parameters are constrained such that only the first latent dimension is discriminant and is a \emph{calibrated} LLR~\cite{brummer2006application,leeuwen13_interspeech}. In this way, the LLR can be handled and set to zero for privacy. \replaced{While, the proposed method can be extended to any binary attributes, t}{T}his paper presents how it can be applied to remove the speaker's sex information into VoxCeleb2~\cite{voxceleb2} speaker embeddings. \added{Indeed, to avoid being a victim of sexism or for any personal reason, users may want to hide their sex.} Next section presents how Bayesian decision theory gives clue for perfect privacy. The third section explains the proposed evidence disentanglement system and the attribute protection strategy. Then, results of its application on the concealment of the sex attribute in speaker embeddings are given in section four.

\section{Bayesian decision \& perfect privacy}
\label{sec:bayesdec}
Considering a set of classes $\mathcal{C} = \{c_0, c_1\}$ and an attacker who wants to infer the class of an observation $x$, the posterior probabilities are given by:
\begin{equation}
    \logit P(c_i | x) = \log \frac{P(x|c_i)}{P(x|c_{\neg i})} + \logit P(c_i),
    \label{eq:bayes}
\end{equation}
where $i\in \{0,1\}$ and $P(c_i)$ is the attacker's prior. \emph{Perfect privacy}~\cite{Nautsch2020}, also known as Shannon's \emph{perfect secrecy}~\cite{shannon}, is reached when the attacker's posterior knowledge remains its prior one. From Equation~\ref{eq:bayes}, one can see that this is achieved when the LLR is zero for all classes and observations. Therefore, in order to remove the information related to a binary attribute in some data, one would like to set the LLR to zero for all samples. This would remove any strength-of-evidence in the data making it attribute-independent~\cite{shannon}. In order to do that, we propose to map the observation's feature vector into a vector space where the first dimension is the LLR (i.e. the \deleted{strength-of-}evidence) and the other dimensions contain remaining variability, called residual, which is independent of the attribute. Because this mapping is invertible, the evidence and residual can be mapped back into the original data's space. Next section presents a first solution that tends toward such mapping.

\section{Disentangling the evidence\\ from the observation}
\label{sec:disent}

In this section, we formalise the disentanglement problem and propose a first solution. Let $\mathcal{X}$ be a $n$-dimensional feature space of some observed data. Assuming that each sample has an \deleted{strength-of-}evidence expressed by a LLR $l \in \mathcal{L} \subset \mathbb{R}$ and has some residual $r=(r_1,\dots r_{n-1})^T$ in $\mathcal{R} \subset \mathbb{R}^{n-1}$, let's define a latent space $\mathcal{Z} = \mathcal{L}\oplus\mathcal{R}$. Our aim is to find, assuming it exists, an invertible mapping $f$ between $\mathcal{X}$ and $\mathcal{Z}$ such that $z=f(x)$ and $x=f^{-1}(z)$ where $x\in \mathcal{X}$ and $z = (l,r_1,\dots r_{n-1})^T \in \mathcal{Z}$. In other words, $f$ disentangles the observation into its \deleted{strength-of-}evidence, and its residual i.e. everything that is independent of $\mathcal{C}$.

\subsection{Class-conditional densities in the latent space}

Due to the \emph{idempotence} property of likelihood-ratio~\cite{meester_slooten_2021}, if one of the class-conditional density of the LLR is Gaussian with mean $\mu$, the other is also Gaussian with an opposite mean $-\mu$ and the same variance $\sigma^2=2\mu$
~\cite{leeuwen13_interspeech,turingstat}. 
For the first component of $z$ to be a LLR, the class-conditional densities in the latent space must respect this property. Therefore, let's define them as follow:
\begin{align}
    z|c_0 &\sim \mathcal{N}(\mu e_1, \Sigma),\\
    z|c_1 &\sim \mathcal{N}(-\mu e_1, \Sigma),
\end{align}
where $\mu \in \mathbb{R}^+$, $e_1=(1,0,\dots 0)^T$ and $\Sigma=\diag(2\mu,1,\dots1)$. It can be shown that in this way, the LLR is given by the first component $z_0 = e_{1}^Tz$ and the idempotence property is respected~\cite{leeuwen13_interspeech}.

\subsection{Invertible mapping between features space and latent space}

Normalizing flow allows latent variable inference and data generation by using an invertible mapping $g$, between the data space and the latent space, learned by data likelihood maximisation. 

\noindent
Let $\mathcal{D} = \{(x^{(0)}, c^{(0)}),\dots (x^{(N-1)}, c^{(N-1)})\}$ be a set of observed samples $x \in \mathcal{X}$ with the labels $c \in \mathcal{C}$. The log-likelihood of the observed data is:
\begin{equation}
\log p_{X|\theta_g,\mu}(\mathcal{D}) =  \sum_{i=0}^{1} \left( \sum_{(x,c) \in \mathcal{D} | c = c_i} \hspace{-0.5cm}\log p_{X|c_i,\theta_g,\mu}(x) \right),
\end{equation}
where $\theta_g$ are the parameters of $g$ and because the mapping is bijective and is applied equally on samples coming from both $c_0$ and $c_1$, the change of variable formula gives:
\begin{equation}
        p_{X|c_i,\theta_g,\mu}(x) = p_{Z|c_i,\mu}(z) \left| \det \left( \frac{\partial g(z)}{\partial z} \right) \right|^{-1},~\forall i \in \{0,1\},
\end{equation}
where $x=g(z)$.
Usually, $g$ is a neural network that can be easily inverted and where the Jacobian determinant (in the formula of variable substitution) can be quickly computed for optimisation~\cite{normflowkobyzev,dinh2017}.

\subsection{Latent distributions parameter and optimisation}
The only parameter of the latent distribution is $\mu$. This parameter appears only on the first component of the latent variable. A straightforward computation shows that, given a batch $\mathcal{B}_Z$ of samples in the latent space, a maximum likelihood estimator of $\mu$ is given by:
\begin{equation}
    \hat{\mu}_{\text{MLE}}(\mathcal{B}_Z) = -1+\sqrt{1+\frac{1}{|\mathcal{B}_Z|}\sum_{z \in \mathcal{B}_Z} (e_{1}^{T}z)^2},
\end{equation}
where $e_{1}^{T}z$ is the first component of the vector $z$.
Like the expectation maximisation algorithm, we propose to do a two-stage iterative optimisation
. We summarise this optimisation strategy in the following where $\alpha$ is the adaptation parameter for the update of $\mu$:
\begin{algorithm}[h]
Choose $\alpha \in [0,1]$,\\
Initialise $\theta_g$ and $\mu$,\\
\For{all batches $\mathcal{B}_X$}{
$\mathcal{B}_Z\leftarrow g^{-1}(\mathcal{B}_X)$\\
$\displaystyle \theta_g \leftarrow \argmax_{\theta_g} \log p_{X| \theta_g, \mu}(\mathcal{B}_X)$\\
$\mu \leftarrow \alpha \mu + (1-\alpha)\hat{\mu}_{\text{MLE}}(\mathcal{B}_Z)$.\\
}
\end{algorithm}
\subsection{Strength-of-evidence manipulation for privacy}
This disentanglement of the LLR in an observation from the irrelevant information allows the manipulation of the data \emph{strength-of-evidence}.
Changing the first component of $z$ permits to change the strength-of-evidence of $x$. Indeed both LLR are equal because the same bijective mapping is applied on samples from both classes:
\begin{equation}
    \frac{p_{Z|c_0}(z)}{p_{Z|c_1}(z)} = \frac{ p_{X|c_0}(x) \left| \det \left( \frac{\partial g(z)}{\partial z} \right) \right| }{ p_{X|c_1}(x) \left| \det \left( \frac{\partial g(z)}{\partial z} \right) \right|} = \frac{ p_{X|c_0}(x)}{ p_{X|c_1}(x)}.
\end{equation}
This can be used in order to make zero-evidence for perfect privacy. Indeed, the LLR can be set to zero following these steps: (1) Map features vector $x$ into the latent space using $z = g^{-1}(x)$, (2) Set the first dimension of $z$ to zero, (3) Map back into the original features space using $x_{\llr_0} = g(z_{\llr_0})$ where $ z_{\llr_0}= (0,r_1,\dots r_{n-1})^T$. By setting the first dimension of the latent variable to zero, this approach allows to alleviate the strength-of-evidence of the data. In the next section, we show a practical application of this approach on attribute-driven privacy.

\section{Application: Attribute-driven privacy}

From now on, we assume that after optimisation, $g^{-1}$ get sufficiently close to the ideal mapping $f$. We will therefore no longer distinguish them. In this section, we present how the proposed approach can be applied to binary attribute-driven privacy. As a toy example, we want to remove the speaker's sex information from embeddings used in speaker verification. As in~\cite{noe21_interspeech}, the user may want to use authentication-by-voice while not disclosing its sex to the service provider. Therefore, before releasing to the remote service the extracted speaker embedding, a sex protection system can be applied on it. We use here kaldi's x-vector~\cite{snyder2018x} as speaker embedding.

\subsection{Experiment}
\label{sec:exp}

In our experiment, we use the same training and testing sets as in~\cite{noe21_interspeech}. V2D and V2T are respectively subsets of VoxCeleb2~\cite{voxceleb2} development and test part on which we respectively train and test our protection system. However, for the protection assessment in~\cite{noe21_interspeech}, the $\Cllrmin$~\cite{brummer2006application} and the ZEBRA's metrics~\cite{Nautsch2020,NOE2022101299} were computed from scores obtained by a state-of-the-art sex classifier trained on non-protected data. With hindsight, we found that this way of assessing the protection can be misleading. Indeed, it only informs if we can fool this classifier and not if the attribute information remains in the protected embeddings. Therefore, these metrics will be here computed on scores obtained by a classifier trained on protected data. To do so, the set V2T (that was not used to train the protection system), is split into a training V2T-train and testing V2T-test part. 46 male and 25 female speakers are randomly chosen to build V2T-train and the remaining speakers i.e. 35 males and 14 females form V2T-test. These two sets contain respectively 17735 and 13944 utterance x-vectors. The classifier will be thus trained and tested respectively on V2T-train and V2T-test. We also provide as in~\cite{noe21_interspeech} a mutual information (MI) measure between the embedding's dimensions and the sex class variable~\cite{ross2014mutual,scikit-learn}.\\

\noindent
\textbf{Proposed system:}\\
The protection system\footnote{Code and models are available: \url{https://github.com/LIAvignon/bridge-features-evidence}} we propose is based on the disentanglement and the protection strategy presented in Section~\ref{sec:disent}. Here, the NF architecture is the Real NVP~\cite{dinh2017} with 6 stacked coupling layers where the scale and translation functions are multilayer perceptrons. The scale function is made of 3 linear layers with 2 LeakyReLU activations and an output Tanh activation. The translation function is made of 3 linear layers with also 2 LeakyReLU activations but no output activation. Adam algorithm with a $10^{-4}$ learning rate is used for optimisation. All $\mu$ initialisations we tried converge to a value close to 9, except for the initialisations really close to zero. The results reported here are those for a $\mu$ initialised at 10 with $\alpha = 0.99$.\\

\noindent
\textbf{Baselines:}\\ We compare the proposed approach to two baselines. The first one is based on LDA and can be expressed with the following whitening:
\begin{equation}
    x \leftarrow x - \frac{ww^T}{||w||^2}x + \frac{1}{2||w||^2}(\mu_{F}^{T}\Sigma_W^{-1} \mu_{F} - \mu_{M}^{T}\Sigma_W^{-1} \mu_{M})w,
\end{equation}
where the class-conditional densities in $\mathcal{X}$ are assumed to be multivariate normal distributions with means $\mu_F$ and $\mu_M$, shared covariance $\Sigma_{W}$, between covariance matrix $\Sigma_{B}$ and $w=\Sigma_{W}^{-1}(\mu_F - \mu_M)$ is the non-zero eigenvalue eigenvector of $\Sigma_{W}^{-1}\Sigma_{B}$. This whitening is like setting the discriminant component such that the estimated LLR is set to zero.
The second baseline called adv-AE were proposed in~\cite{noe21_interspeech}. It is based on adversarial disentanglement autoencoding. An encoder-decoder and a sex-classifier on the encoded representation are trained in an adversarial manner. The encoder tries to fool the classifier while the latter tries to predict the right sex. This would reduce the sex information in the encoded representation. The decoder takes an additional input: a score related to the sex information. During testing, this score is set to a constant $0.5$ in order to reduce the sex information.

\subsection{Results}
\begin{figure}[H]
        \centering
        \begin{subfigure}[b]{0.23\textwidth}
            \centering \includegraphics[height=0.98\textwidth]{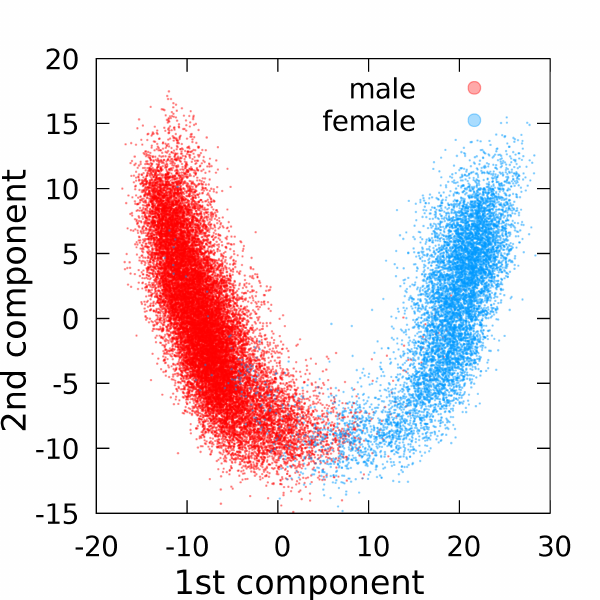}
            \caption{{\small Principal components of $x$}}    
        \end{subfigure}
        \hfill
        \begin{subfigure}[b]{0.23\textwidth}  
            \centering 
            \includegraphics[height=0.98\textwidth]{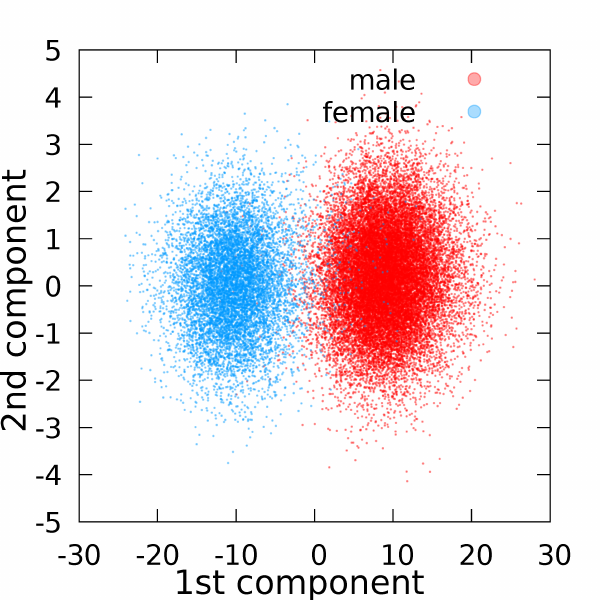}
            \caption{{\small Two first components of $z$}}    
        \end{subfigure}
        \caption{V2T's x-vectors visualisation in the original and latent space. In the latent space (b), class-conditional densities look Gaussian and only the first component is discriminant while this is not the case in the original feature space (a).}
        \label{fig:visulatent}
\end{figure}

Before discussing the protection and the utility results, let's visualise in Figure~\ref{fig:visulatent} how the data looks in the latent space. One can see that the mapping makes the two class-conditional densities Gaussian like and only the first component is discriminant as expected.

\noindent
\textbf{Protection ability assessment:}\\
A 2-layers perceptron sex classifier is trained on four versions of V2T-train: without protection, with the LDA and adv-AE protection and with the proposed method that we call NFzLLR. Each classifier is tested on the corresponding V2T-test set. Results are shown in Table~\ref{tab:classif_results}. Even if the classifier learned to separate relatively well the sex on the training data due to overfitting, with the proposed method, it hardly generalises to the test set in comparison to the baselines. The best baseline is the adv-AE approach because its $\Cllrmin$ on V2T-test is larger than for LDA and the expected amount of sex information disclosed by scores ($D_{ECE}$) is lower. However, the proposed approach outperforms both baselines in terms of $\Cllrmin$ and $\Dece$.

\begin{table}[ht]
    \centering
    \caption{Sex classification performance on non-protected and protected V2T. The first line is for non-protected data, the second and third lines are for the two baselines and the last line is for the proposed system. For good privacy, $\Cllrmin$ has to be close to 1 and $\Dece$ should be as low as possible on V2T-test. Based on that, the proposed system provides better protection compared to the baselines.}
    \label{tab:classif_results}
    \begin{tabular}{ccc|cc}
    
            \cline{2-5}
            & \multicolumn{2}{c}{\makecell{V2T-train}}
            & \multicolumn{2}{c}{\makecell{V2T-test}}
            \\
            \cline{2-5}

             & $\Cllrmin${\tiny $10^{-2}$} & $\Dece$ & $\Cllrmin${\tiny $10^{-2}$} & $\Dece$\\
            \hline
            original data & 8.39 & 0.658 & 2.12 & 0.703 \\
            
            LDA & 12.20 & 0.628 & 57.75 & 0.295 \\
    
            adv-AE & 30.43 & 0.493 & 74.21 & 0.179 \\
            
            \textbf{NFzLLR} & \textbf{48.45} & \textbf{0.362} & \textbf{95.75} & \textbf{0.029} \\
            \hline
    \end{tabular}
\end{table}
\begin{table}[ht]
    \centering
    \caption{Absolute number of samples per tag on V2T-test (relative values obscure extrema). Bold numbers indicate the worst-case tag.}
    \begin{tabular}{c|c|c|c|c|c|c}
    & A & B & C & D & E & F  \\
    \hline
         original data & 15 & 194 & \textbf{13735} & 0 & 0 & 0 \\
         LDA & 11592 & 844 & \textbf{1508} & 0 & 0 & 0 \\
         adv-AE & 12454 & 684 & \textbf{806} & 0 & 0 & 0 \\
         \textbf{NFzLLR} & 13821 & \textbf{123} & 0 & 0 & 0 & 0 \\
         \hline
    \end{tabular}
    \label{tab:absnumbertag}
\end{table}
\noindent
\newpage
\noindent
In~\cite{Nautsch2020}, ZEBRA's metrics release also the score with the highest strength-of-evidence corresponding to the worst-case scenario. Along with it, a categorical tag is given: the latter is a letter from A to F, from better to worst. Table~\ref{tab:absnumbertag} gives the number of samples per tag for V2T-test. One can see the worst-case tag for the proposed system is B with a low amount of scores compared to the other lines with a worst-case tag C.
Mutual information measures are given in Table~\ref{tab:mi}. It confirms the better protection ability of the proposed system. On V2T the drop of MI is about 96.47\% while for the LDA and the adversarial approach, the decrease is about 88.00\% and 90.00\%.
\begin{table}[ht]
        \centering
        \caption{Mutual information measures between the x-vectors and the sex class variable $y$ on V2D and V2T.}
    \label{tab:mi}
        \begin{tabular}{ccc}
            \cline{2-3}
            & \multicolumn{2}{c}{MI {\tiny $10^{-2}$} [bit per dimension]}\\
            \hline
             &  V2D  & V2T \\
            \hline
            original data & 18.7 & 19.0 \\
            LDA & 1.43 & 2.28 \\
            adv-AE & 1.0 &  1.90 \\
            \textbf{NFzLLR} & \textbf{0.14} & \textbf{0.67}\\
            \hline
        \end{tabular}
\end{table}

\begin{figure}[h]
        \centering
        \begin{subfigure}[b]{0.23\textwidth}
            \centering 
            \includegraphics[height=1.1\textwidth]{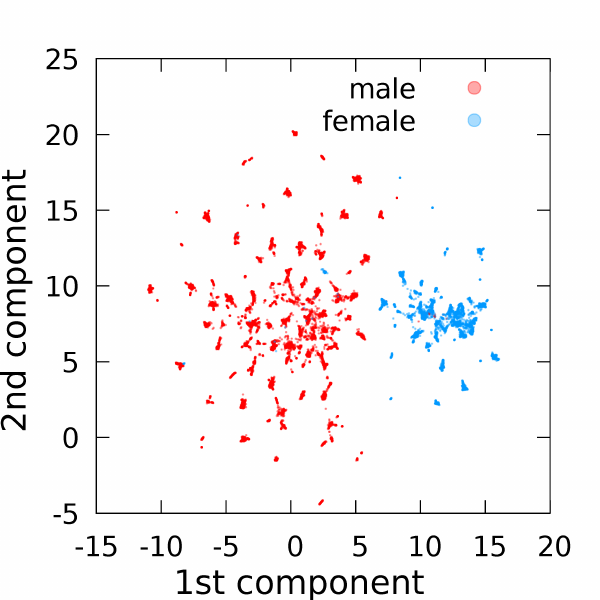}
            \caption{{\small Original}}    
        \end{subfigure}
        \hfill
        \begin{subfigure}[b]{0.23\textwidth}  
            \centering 
            \includegraphics[height=1.1\textwidth]{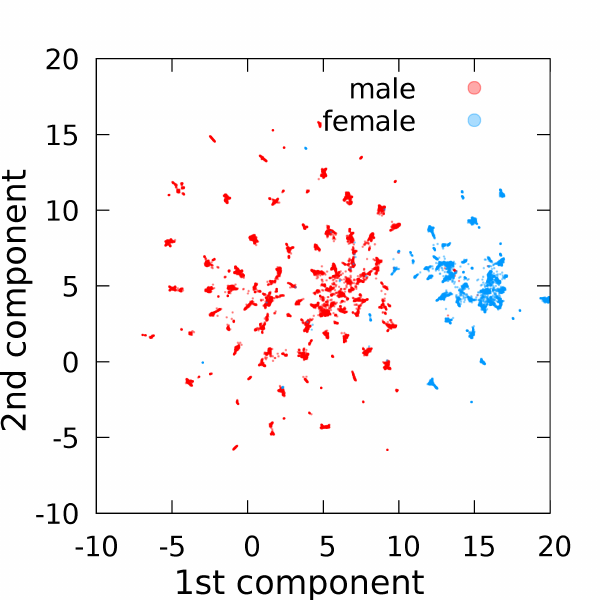}
            \caption{{\small LDA}}    
        \end{subfigure}
        \vskip\baselineskip
        \vspace{-0.4cm}
        \begin{subfigure}[b]{0.23\textwidth}   
            \centering
            \includegraphics[height=1.1\textwidth]{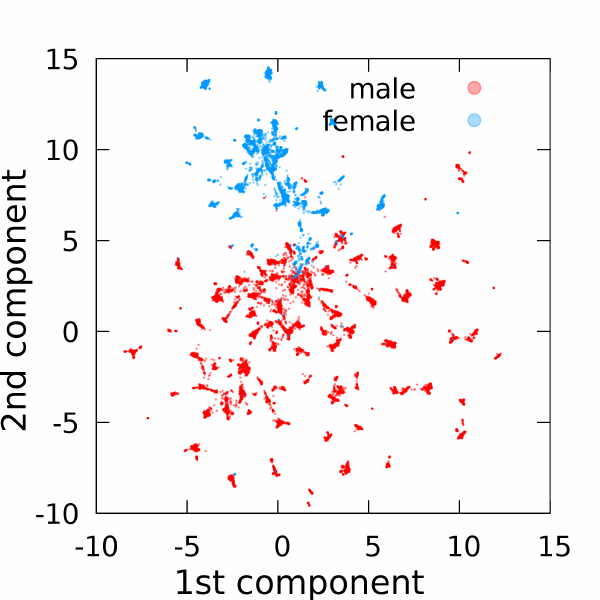}
            \caption{{\small adv-AE}}    
        \end{subfigure}
        \hfill
        \begin{subfigure}[b]{0.23\textwidth}   
            \centering 
            \includegraphics[height=1.1\textwidth]{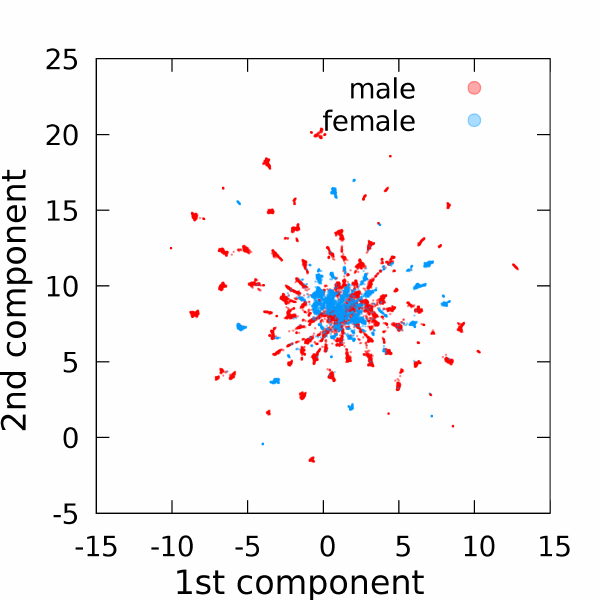}
            \caption{{\small NFzLLR}}    
        \end{subfigure}
        \caption{UMAP visualisation (with euclidean metric and 30 neighbors)~\cite{mcinnes2018umap} of the original and protected V2T's x-vectors. Even with the LDA and adversarial based methods, unsupervised visualisation allows separation of the sex while with the proposed approach, male and female x-vectors are mixed.}
        \label{fig:umap}
\end{figure}

\noindent
The better protection given by the proposed approach is also shown in Figure~\ref{fig:umap}. It provides 2D visualisations of the embeddings using the UMAP dimensionality reduction and visualisation~\cite{mcinnes2018umap}. We can see that, even with the LDA and adv-AE methods, the unsupervised visualisation allows separation of the sex while with the new approach, male and female x-vectors are mixed.

\noindent
In addition of making sure that the attribute has been hidden, we want to see if the utility is preserved i.e. if we can still do automatic speaker verification.
\begin{table}[ht]
    \centering
    \caption{Speaker verification performance. Even if the proposed system slightly alter the ASV performance, it does better than the adversarial-autoencoding system.}
    \label{tab:asvres}
    \begin{tabular}{ccc}
        \hline
         & EER [\%] & $\Cllrmin$ \\
        \hline
        original data & 1.72 & 0.067 \\
        adv-AE & 2.36 & 0.097 \\
         \textbf{NFzLLR} & \textbf{2.11} & \textbf{0.086} \\
        \hline
    \end{tabular}
\end{table}

\noindent
\textbf{Automatic speaker verification results:}\\
We use the same speaker verification protocol as in~\cite{noe21_interspeech}. The common probabilistic linear discriminant analysis (PLDA)~\cite{pldaIoffe} is used to compute the scores from the comparisons between enrolment and probe segments. The PLDA is trained on non-protected data. Indeed, we suppose that the authentication side does not know that the user is protecting its data. Therefore, the authentication has a general ASV backend and the user applies protection on both reference and probe segments. The equal-error-rate (EER) and $\Cllrmin$ for the original data, and protected with the adv-AE and the proposed approach are given in Table~\ref{tab:asvres}. Here, metrics are of course given for the classification between \emph{target} and \emph{impostor} trials while in Table~\ref{tab:classif_results}, it was for the classification between \emph{female} and \emph{male} utterances. Because the protection ability of the LDA is not competitive, its results are not reported here.
In addition of having a good protection ability, the method we proposed results in better speaker verification in comparison to the adversarial approach. This difference could be explained by the fact that the adversarial-autoencoder has some reconstruction error contrary to the normalising flow based approach. However, this adversarial-autoencoder's drawback can probably be reduced by investigating more flexible architecture and better tuning.

\section{Conclusion}
This paper proposed a non-linear discriminant analysis that allows to map feature vector into a space where the LLR related to a binary attribute, and the residual i.e.\ everything that is independent of this attribute, are disentangled. Because the mapping is invertible, this approach can be used for attribute-driven privacy by manipulating the strength-of-evidence of the data. More precisely, binary information can be removed from some data by setting the estimated LLRs to zero. This would result in perfect privacy, also known as zero-evidence and perfect secrecy. Our experiments on the VoxCeleb2 speaker embeddings showed the ability of our approach to remove the sex information in the embeddings while preserving fairly well the speaker verification ability. The proposed approach is theoretically justifiable and outperforms, in terms of protection and utility, our previous work based on adversarial disentanglement.

The proposed approach is based on the LLR paradigm and zero-evidence which are well established for binary problems. However, most of the speaker attributes, like the nationality or the age, are not binary, even not discrete. Therefore, our future works will focus on the generalisation of our method and the zero-evidence concept to attributes that consist of more than two classes.

\section{Acknowledgements}
This work was supported by the VoicePersonae project ANR-18-JSTS-0001.

\newpage
\bibliographystyle{IEEEbib}
\balance\bibliography{bib}

\end{document}